\documentclass[runningheads]{llncs}
\usepackage{url}            
\usepackage{booktabs}       
\usepackage{amsfonts}       
\usepackage{nicefrac}       
\usepackage{microtype}      
\usepackage{amsmath}
\usepackage{graphicx} 
\usepackage[table,xcdraw]{xcolor} 
\usepackage{colortbl} 
\usepackage{xcolor}
\usepackage{marvosym}
\usepackage{hyperref}
\usepackage{amssymb}
\usepackage{graphicx,verbatim}
\usepackage{graphicx,verbatim}
\usepackage{multirow}
\definecolor{electricindigo}{rgb}{0.44, 0.0, 1.0}
\definecolor{lightblue}{RGB}{240,245,255}
\definecolor{darkblue}{RGB}{40,40,85}
\definecolor{babyblue}{rgb}{0.54, 0.81, 0.94}
\definecolor{pearDark}{HTML}{2980B9}
\definecolor{pearDarker}{HTML}{1D2DEC}
\usepackage[capitalize]{cleveref}
\crefname{section}{Sec.}{Secs.}
\Crefname{section}{Section}{Sections}
\Crefname{table}{Table}{Tables}
\crefname{table}{Tab.}{Tabs.}
\usepackage{fontawesome5}
\usepackage[T1]{fontenc}
\definecolor{electricindigo}{rgb}{0.44, 0.0, 1.0}
\definecolor{lightblue}{RGB}{240,245,255}
\definecolor{darkblue}{RGB}{40,40,85}
\definecolor{babyblue}{rgb}{0.54, 0.81, 0.94}
\definecolor{pearDark}{HTML}{2980B9}
\definecolor{pearDarker}{HTML}{1D2DEC}
\usepackage[capitalize]{cleveref}
\crefname{section}{Sec.}{Secs.}
\Crefname{section}{Section}{Sections}
\Crefname{table}{Table}{Tables}
\crefname{table}{Tab.}{Tabs.}
\usepackage{fontawesome5}

\definecolor{deblue}{RGB}{11,132,147}
\definecolor{ocra}{RGB}{204, 119, 34}

\usepackage{tikz}
\newcommand{\fcircle}[2][red,fill=red]{\tikz[baseline=-0.5ex]\draw[#1,radius=#2] (0,0.03) circle ;}

\usepackage{multirow}
\usepackage{colortbl}
\usepackage{tabulary}
\usepackage{etoolbox}
\usepackage{tikz}
\usepackage{pifont}
\makeatletter
\@namedef{ver@everyshi.sty}{}
\makeatother
\usepackage{tikz}
\usepackage{pifont}
\usepackage{graphicx,verbatim}
\begin{document}
\title{Implicit U-KAN2.0: Dynamic,  Efficient and Interpretable Medical Image Segmentation}
\titlerunning{Implicit U-KAN2.0}
%
\author
{
Chun-Wun Cheng\inst{1}\textsuperscript{*}
\and
Yining Zhao\inst{2}\textsuperscript{*} 
\and
Yanqi Cheng  \inst{1} 
\and
Javier A. Montoya-Zegarra \inst{3,4,5}   
\and 
Carola-Bibiane Schönlieb  \inst{1}
\and
Angelica I Aviles-Rivero\inst{6}\textsuperscript{(\Letter)}}
%
\authorrunning{CW Cheng et al.}
%
\institute{
Department of Applied Mathematics and Theoretical Physics, University of Cambridge, UK \and
Siebel School of Computing and Data Science, University of Illinois Urbana-Champaign, USA 
\and
Zurich University of Applied Sciences, Switzerland
\and
Lucerne Cantonal Hospital, Switzerland
\and
Lucerne University of Applied Sciences and Arts, Switzerland
\and
Yau Mathematical Sciences Center, Tsinghua University, China \{\email{aviles-rivero@tsinghua.edu.cn}\}
}
\maketitle              
\renewcommand{\thefootnote}{}
\footnotetext{\textsuperscript{$*$} Equal contributions.}
\begin{abstract}
Image segmentation is a fundamental task in both image analysis and medical applications. State-of-the-art methods predominantly rely on encoder-decoder architectures with a U-shaped design, commonly referred to as U-Net. Recent advancements integrating transformers and MLPs improve performance but still face key limitations, such as poor interpretability, difficulty handling intrinsic noise, and constrained expressiveness due to discrete layer structures, often lacking a solid theoretical foundation.In this work, we introduce Implicit U-KAN 2.0, a novel U-Net variant that adopts a two-phase encoder-decoder structure. In the SONO phase, we use a second-order neural ordinary differential equation (NODEs), called the SONO block, for a more efficient, expressive, and theoretically grounded modeling approach. In the SONO-MultiKAN phase, we integrate the second-order NODEs and MultiKAN layer as the core computational block to enhance interpretability and representation power. Our contributions are threefold. First, U-KAN 2.0 is an implicit deep neural network incorporating MultiKAN and second order NODEs, improving interpretability and performance while reducing computational costs. Second, we provide a theoretical analysis demonstrating that the approximation ability of the MultiKAN block is independent of the input dimension. Third, we conduct extensive experiments on a variety of 2D and a single 3D dataset, demonstrating that our model consistently outperforms existing segmentation networks. Project Website: \href{https://math-ml-x.github.io/IUKAN2/}{https://math-ml-x.github.io/IUKAN2/}


\end{abstract}

\section{Introduction}
Image segmentation is crucial in many applications, particularly in medical imaging and computational sciences, where accurate delineation of anatomical structures is essential. Traditionally, this task has relied on manual annotation by clinicians—a process that is both time-consuming and costly. Recently, deep learning methods, especially convolutional neural networks (CNNs), have significantly improved segmentation accuracy. Among these approaches, U-Net~\cite{ronneberger2015u} and its variants have become foundational due to their effective encoder-decoder architectures with skip connections. Building on U-Net’s success, numerous extensions have been developed to enhance performance further. For instance, ResUNet~\cite{zhang2018road} incorporated residual units to facilitate the training of deeper networks. Inspired by this work, various block-based architectures~\cite{zhao2017pyramid,li2018h,zhang2018bi} have also been explored. 
CNN-based models excel at capturing local dependencies but often struggle with global contextual information. To address this, researchers have integrated Transformers into U-Net architectures, with Attention U-Ne~\cite{oktay2018attention} pioneering the use of attention mechanisms. Later, transformer-based U-Net variants~\cite{petit2021u,cao2022swin} were introduced to reduce computational costs while maintaining performance. The Mamba framework \cite{gu2023mamba} emerged as an alternative to traditional Transformers, offering comparable performance with reduced computational complexity from quadratic to linear. Models like U-Mamba \cite{ma2024u} and Swin-UMamba \cite{liu2024swin} further build on the Mamba framework, incorporating aspects of nnU-Net \cite{isensee2018nnu} architecture, which requires pre-training.
Both CNN-based and transformer-based models discretise continuous functions, while Continuous U-Net \cite{cheng2023continuous} offers a continuous block to address this. The continuous formulation of Second Order NODEs \cite{norcliffe2020second,chen2018neural} enables $\mathcal{O}(1)$ memory cost and has been applied in various tasks \cite{zhang2024cross,zhang2024node,ordonezmissing}. Inspired by Kolmogorov–Arnold Networks (KANs) \cite{liu2024kan}, which use learnable activation functions at edges to optimise feature representation, U-KAN \cite{li2024u} integrates a Tokenised KAN Block with a Convolution Block in U-Net, but relies only on addition. MultiKAN \cite{liu2024kan} extends KANs by incorporating both addition and multiplication, improving capacity and interpretability.\\
\indent In this study, we address the limitations of U-KAN by introducing \textbf{Implicit U-KAN2.0}, which builds on implicit neural networks for improved efficiency and interpretability. \fcircle[fill=deblue]{2pt} \textbf{Our approach introduces two key novelties:} 1) the \textit{Second Order Neural ODE (SONO) block}, which transforms discrete functions into continuous ones while maintaining constant memory cost, and 2) the \textit{SONO-MultiKAN block}, which integrates SONO with a tokenised MultiKAN layer for enhanced representation power.
Compared to U-KAN, our method \textbf{redesigns}  the architecture by replacing convolutional blocks with SONO, improving discretisation and stability. Furthermore, we introduce a \textit{bottleneck module} to refine the flow of information between the encoder and decoder, optimising feature retention and improving overall model performance.
Instead of additive skip connections, we employ feature concatenation to preserve richer representations. Additionally, unlike KAN-based U-Nets, which lack full GPU compatibility, our model fully optimised for GPU-based training while maintaining constant memory costs and ensuring scalability.
\fcircle[fill=deblue]{2pt} \textbf{Our contributions:} I) We introduce a novel implicit deep neural network powered by our SONO block, which  enhances the model’s ability to evolve features continuously, improving both accuracy and stability in segmentation tasks.
II) We provide theoretical analysis demonstrating that the MultiKAN block’s approximation, with high expressiveness, is independent of input dimensionality. III) Extensive experiments on multiple 2D and 3D datasets confirm that our model outperforms existing segmentation networks.

\section{Methodology}
\begin{figure}[t!]
\centering
\includegraphics[width=1\textwidth]{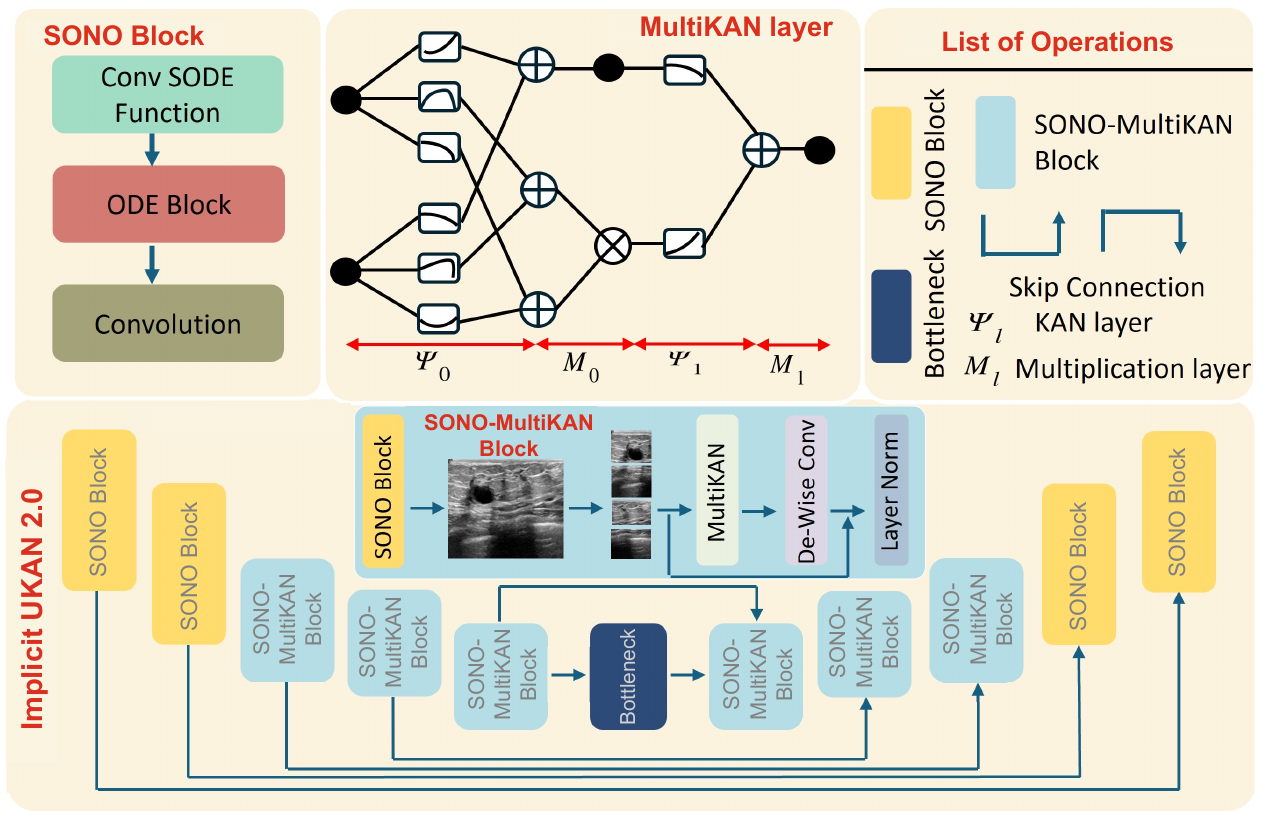}
\caption{Overview of Implicit U-KAN2.0. The upper section shows the SONO Block and the MultiKAN layer, while the lower section provides the overall architecture.}
\label{fig::teaser}
\end{figure}

This section outlines the key contributions of Implicit U-KAN2.0, highlighting its dynamic feature evolution and enhanced interpretability, which result in higher accuracy and greater efficiency in medical image segmentation tasks. Additionally, we provide a detailed overview of its architecture. 

\subsection{Implicit U-KAN2.0 is Dynamic}

The first key highlight of our Implicit U-KAN2.0 is using Second-Order Neural Ordinary Differential Equations (NODEs) to model the continuous evolution of feature representations, namely SONO block. This dynamic modelling allows for smoother learning trajectories and faster convergence compared to existing methods. The second-order NODEs governing feature evolution is:
$\bold{\ddot{x}}(t) = f(\bold{x}, \bold{\dot{x}}, t, \theta_f), \quad \bold{\dot{x}}(t_0) = g(x_0, \theta_g), \quad  \bold{x}(t_0) = x_0$.
where $f$ and $g$ are both neural networks and $\bold{x}(t)$ represents the feature vector. The feature was evaluated as a continuous representation and solved by the ode solver in both forward and backward propagation. In addition, the adjoint method was introduced during backpropagation in order to achieve a constant memory cost.
By incorporating velocity into the system and extending the solution space to three dimensions, Implicit U-KAN 2.0 achieves faster convergence to optimal feature representations, enhanced stability, and more precise segmentation boundaries. This is particularly important medical imaging
where precise boundary delineation is required. To achieve this, we introduce the velocity term \( \mathbf{v}(t) = \mathbf{x}'(t) \), reformulating the problem into a system of first-order ordinary differential equations (ODEs):$\bold{\dot{x}}(t) = \mathbf{v}(t),  \hspace{0.5cm} \bold{\dot{v}}(t) = f(\mathbf{v}(t), t, \boldsymbol{\theta})$. This transformation expands the phase space to \( [\mathbf{x}(t), \mathbf{v}(t)]^\top \in \mathbb{R}^{2n} \), which accelerates convergence by enabling trajectory corrections in both position and velocity enabling trajectory corrections in both position and velocity. As a result, convergence is accelerated while maintaining stability. Additionally, the RK4 method is employed to approximate a more stable solution, further improving numerical robustness.
This continuous feature evolution ensures smoother transitions in learned representations, which is crucial for medical image segmentation. By capturing gradual changes, it reduces overfitting and makes the network more resilient to noisy, common in medical imaging. This leads to more stable feature extraction and improved segmentation accuracy, even with low-quality images.

\subsection{Implicit U-KAN2.0 with higher interpretability}
The second key contribution of Implicit U-KAN2.0 involves the integration of Second-order NODEs and MultiKAN layer. Similarly to a SONO block, we first transform the discrete feature vector into a continuous second-order function. Then, the features are further refined using tokenlised MultiKAN layer. In KAN, weight matrix was learned by a b-spline function. Inspired by the Kolmogorov-Arnold representation theorem (KART), any continuous function defined in a high-dimensional space can be expressed as a finite combination of continuous functions, each depending on a single variable, along with summation operations. Formally, for any smooth function $f$,
$f(\mathbf{x}) = \sum_{q=1}^{2n+1} \Phi_q\left( \sum_{p=1}^n \varphi_{q,p}(x_p) \right)$,
where \( \mathbf{x} \) is the input feature vector, \( \varphi_{q,p} \) and \( \Phi_q \) are univariate functions capturing feature interactions. KAN with 
$L$ layers can be written as a set of composition functions:
$\mathrm{KAN}(x) = \left( \Phi_{L-1} \circ \cdots \circ \Phi_1 \circ \Phi_0 \right)(x)$

However, to enhance both interpretability and capacity, MultiKAN was introduced such that incorporates both addition and multiplication operations. MultiKAN extends the standard KANs by interleaving multiplication sub-layers with the conventional addition-based layers.
Concretely, the network is defined by two sequences of array sizes: 
$m^{a} = [m^{a}_0, m^{a}_1, \dots, m^{a}_{l}]$
for addition arrays, and $m^{n} = [m^{n}_0, m^{n}_1, \dots, m^{n}_l]$ for multiplication arrays. A multiplication layer can be divided into two components. The first component transforms the addition nodes without performing any additional actions, while the second component carries out the multiplication. A single MultiKAN layer $\Psi_i$ is constructed by combining the corresponding KAN layer $\Phi_i$ with  $M_i$, and the full MultiKAN is obtained by composing these combined layers.
\begin{equation}
\text{MultiKAN}(x) 
\;=\; \bigl(\Psi_L \circ \Psi_{L-1} \circ \dots \circ \Psi_1\bigr)(x).
\label{multikan}
\end{equation}
This architecture enriches the interactions among feature arrays and can leverage standard training and regularization methods from the KAN framework because we do not perform additional training in this sub-layer.
Similar to KANs, we demonstrated that MultiKAN's approximation ability is not influenced by dimensionality; rather, it depends on the residual rate.

\begin{theorem} [Kolmogorov Arnold theorem for MultiKAN]
Suppose that we define $\text{MultiKAN}(x) = f$ as in~\eqref{multikan} and assume $\Psi_{l,i,j}$ is (k+1)-times continuously differentiable. Then, for a given function $G$-grid B-spline functions, there exist exists a constant $C$, dependent on $f$, such that the following holds for all integers $m$ with  $0 \le m \le k$:
$\bigl\|\,\text{MultiKAN}(x) - \bigl(\Phi_{L-1}^G \circ \Phi_{L-2}^G \circ \cdots \circ \Phi_1^G \circ \Phi_0^G\bigr)x \bigr\|_{C^m}
\;\le\; C\, G^{-k-1+m}$.

\end{theorem}

\begin{proof}
This proof follows a similar approach to that in Theorem
2.1~\cite{liu2024kan} and we provide a sketch of the proof in here. By the B-spline lemma and  the continuity of the functions $\Psi_{l,i,j}$, these functions remain uniformly bounded on any domain. Then  we can write 
$\text{MultiKAN}(x) - \bigl(\Phi_{L-1}^G \circ \Phi_{L-2}^G \circ \cdots \circ \Phi_1^G \circ \Phi_0^G\bigr)x = Q_{L-1} + Q_{L-2} + ... + Q_1+ Q_0$,
where $Q_l = \bigl(\Phi_{L-1}^G \circ \cdots \circ  \Phi_{l+1}^G\circ \,\Phi_l\, \circ\Phi_{l-1} \circ \cdots \circ \Phi_0\bigr)x
\;-\;
\bigl(\Phi_{L-1}^G \circ \cdots \circ \Phi_{l+1}^G\circ \,\Phi_l^G\, \circ\Phi_{l-1} \circ \cdots \circ \Phi_0\bigr)x$ was bounded by $CG^{-k-1+m}$.

\end{proof}

\subsection{Implicit UKAN2.0 Architecture}
The architecture of Implicit U-KAN2.0 seamlessly integrates its components within a unified encoder-decoder framework. The operation workflow is delineated into two phases: the SONO Phase and the SONO-MultiKAN Phase. Figure \ref{fig::teaser} shows the overall architecture and the  {detailed components} of each Block.  \\
\textbf{SONO Phase:} The initial input image $X_0$ is first processed to establish an initial velocity. Subsequently, the extracted features are propagated through a SONO Block, yielding an output defined by: $X_L = \text{Conv}(\text{ODEBlock}(X_{L-1}))$, 
where $X_{\ell} \in \mathbb{R}^{H_{\ell} \times W_{\ell} \times C_{\ell}}$, where $X_{\ell} \in \mathbb{R}^{H_{\ell} \times W_{\ell} \times C_{\ell}}$ denotes the feature map at the $l$-th layer. Within the SONO Block, we initially employ a function to model the second-order derivative. The resulting output is then fed into an ODE Block, thereby facilitating the approximation of continuous second-order NODEs. Subsequently, we enhance the number of features and perform downsampling via a convolutional layer.
In the forward pass, the ODE block is formulated as
$\text{ODEBlock}(v_t) = \text{ODESolve}(\mathbf{v}(t_0), f, t_0, t_u, \theta)$
and the loss function is defined by $L(\textbf{v}(t_u)) = L\left( \textbf{v}(t_0) + \int_{t_0}^{t_u} f (\textbf{v}(t), t, \theta) \, dt \right) = L\left(\text{ODESolve}(\textbf{v}(t_0), f, t_0, t_u, \theta )\right)$. 
\\ \indent \textbf{SONO-MultiKAN Phase:}
In the SONO-MultiKAN phase, the ODE Block is initially applied, followed by the tokenization process to prepare the features for the MultiKAN layer. MultiKAN utilizes learnable activation functions to enhance interpretability. First, the SONO output \(X_L\) from the previous two encoders are transformed into a collection of flattened two-dimensional patches, consistent with the methodology described in~\cite{dosovitskiy2020image}. Specifically, \(X_L\) is partitioned into $M$ patches, each of dimensions 
$K \times K$, such that $M = \frac{H_L \times W_L}{K^2}$. Each patch, denoted by $X_L^i \in \mathbb{R}^{k^2 \cdot C_L}$, is subsequently projected into a $d$ -dimensional embedding space through a learnable linear mapping $E \in \mathbb{R}^{(P^2 \cdot C_L) \times d}$. Consequently, the initial token embeddings are represented as $Z_0 = \bigl[X_L^1 E; \; X_L^2 E; \; \dots; \; X_L^N E\bigr]$. { These embeddings are then processed by a MultiKAN module with interleaved multiplicative layers to capture higher-order, non-linear feature interactions.} Next, a three-layer MultiKAN module is employed to further extract features, which is then followed by a depth-wise convolutional layer (De-WiseConv). A residual connection is incorporated by adding the original token embeddings, mathematically expressed as: $Z_K = LN(Z_{k-1} + \text{De-WiseConv}(\text{MultiKAN}((\text{ODEBlock}(X_{k-1}))) = LN(Z_{k-1} + \text{De-WiseConv}(\text{MultiKAN}(Z_{k-1})$. The MultiKAN block provides structural transparency through tokenised basis functions with explicit mathematical roles, unlike saliency maps, which offer approximate explanations for black-box models.  
\\ \indent \textbf{Decoder:} Similarly to UNet architecture, the bottleneck was introduced and serves as the critical bridge between the encoder and the decoder. Additionally, skip connections were incorporated, and the decoder comprises two consecutive dynamic blocks and three consecutive MultiKAN blocks for upsampling.  We use binary cross-entropy loss for the loss function.

\section{Experiments} 
\textbf{ Datasets and Implementation Detail. }We performed experiments using our Implicit U-KAN 2.0 on three distinct 2D medical imaging datasets, namely Kvasir-SEG~\cite{jha2020kvasir}, ISIC Challenge (ISIC)~\cite{gutman2016skin} and Breast Ultrasound Images (BU Images)~\cite{al2020dataset}. These datasets encompass a wide range of data types and exhibit substantial variability in terms of image dimensions, segmentation mask fidelity, and overall dataset size. 
Kvasir-SEG contains 1,000 colonoscopy images (800 for training, 200 for testing) at 256×256 for polyp segmentation. The ISIC Challenge dataset has 1,279 dermoscopy images (990 training, 289 testing) at 512×512 for skin lesion segmentation. The Breast Ultrasound Images dataset includes 647 images (518 training, 129 testing) at 256×256 for breast lesion segmentation. For our 3D experiments, we used the Medical Segmentation Decathlon’s Spleen dataset, which comprises 61 CT volumes (41 training, 20 testing) focusing on the spleen.
We implemented the Implicit U-KAN 2.0 model on an NVIDIA A100 40G GPU for testing on the Kvasir-SEG dataset and ISIC Challenge, with batch sizes of 15 and 3 respectively. For the Breast Ultrasound Images, an RTX 4070 Super GPU was used with a batch size of 4. The model was trained for 500 epochs at a learning rate of 0.0001 using early stopping.

\subsection{Experiment results and Ablation Studies}

\begin{table}[t!]
\renewcommand{\arraystretch}{1.15}
\caption{The comparison of segmentation performance across various methods for three different datasets.}
\label{tab:results-2D}
\begin{tabular}{
    l | l |
    >{\centering\arraybackslash}p{1.15cm} |
    >{\centering\arraybackslash}p{1.15cm} |
    >{\centering\arraybackslash}p{1.15cm} |
    >{\centering\arraybackslash}p{1.15cm} |
    >{\centering\arraybackslash}p{1.15cm} |
    >{\centering\arraybackslash}p{1.15cm} |
    >{\centering\arraybackslash}p{1.15cm} |
    >{\centering\arraybackslash}p{1.15cm}
}
\hline
\rowcolor[HTML]{EFEFEF} 
Dataset & Metric &
\begin{tabular}[c]{@{}l@{}}U-Net \,\end{tabular} &
\begin{tabular}[c]{@{}l@{}}Trans\\ UNet \,\end{tabular} &
\begin{tabular}[c]{@{}l@{}}UNext\end{tabular} &
\begin{tabular}[c]{@{}l@{}}USODE \,\end{tabular} &
\begin{tabular}[c]{@{}l@{}}Rolling\\ -Unet \,\end{tabular} &
\begin{tabular}[c]{@{}l@{}}MLLA\\ -UNet\end{tabular} &
\begin{tabular}[c]{@{}l@{}}U-KAN \,\end{tabular} &
\begin{tabular}[c]{@{}l@{}}Ours\end{tabular} \\
\hline
& Dice $\uparrow$        & 0.7741  & 0.7824  & 0.7583  & 0.7465  & 0.8078  & 0.6962  & 0.7331  & \textbf{\cellcolor[HTML]{D7FFD7} 0.8456} \\
& HD95 $\downarrow$       & 44.75  & 37.97   & 41.12   & 45.20    & 33.74   & 53.62   & 48.40    & \textbf{\cellcolor[HTML]{D7FFD7} 25.26} \\
& Acc. $\uparrow$      & 0.9023  & 0.9064  & 0.8926  & 0.9031  & \textbf{\cellcolor[HTML]{D7FFD7} 0.9278} & 0.8791  & 0.8949  & 0.9134 \\
& IoU $\uparrow$        & 0.4857  & 0.5142  & 0.4657  & 0.4264  & 0.5705  & 0.3574  & 0.4207  & \textbf{\cellcolor[HTML]{D7FFD7} 0.6396} \\
\multirow{-5}{*}{\begin{tabular}[c]{@{}l@{}}Kvasir\\ -SEG\end{tabular}} 
& F1 $\uparrow$         & 0.6074  & 0.6259  & 0.5821  & 0.5506  & 0.6585  & 0.4653  & 0.5289  & \textbf{\cellcolor[HTML]{D7FFD7} 0.7490} \\
\hline
& Dice $\uparrow$       & 0.9091  & 0.9233   & 0.9155  & 0.8998  & 0.9195  & 0.9243 & 0.9190  & \textbf{\cellcolor[HTML]{D7FFD7} 0.9330} \\
& HD95 $\downarrow$      & 15.86   & 10.48   & 14.25   & 24.90   & 9.67    & 10.09   & 12.45  & \textbf{\cellcolor[HTML]{D7FFD7} 7.61}  \\
& Acc. $\uparrow$      & 0.9402  & 0.9505  & 0.9448  & 0.9334  & 0.9497  & 0.9514  &  0.9485 & \textbf{\cellcolor[HTML]{D7FFD7} 0.9577} \\
& IoU $\uparrow$        & 0.8044  &0.8339  &0.8159  &0.7854  & 0.8238  & 0.8350   & 0.8244  & \textbf{\cellcolor[HTML]{D7FFD7} 0.8513} \\
\multirow{-5}{*}{\begin{tabular}[c]{@{}l@{}}ISIC\end{tabular}} 
& F1 $\uparrow$         & 0.8784  &0.9009 & 0.8885  &0.8673  & 0.8934 & 0.9005    & 0.8924  & \textbf{\cellcolor[HTML]{D7FFD7} 0.9128} \\
\hline
& Dice $\uparrow$       & 0.7833  & 0.8117 & 0.7812 & 0.7573  & 0.8187 & 0.7739   & 0.8140    & \textbf{\cellcolor[HTML]{D7FFD7} 0.8397} \\
& HD95 $\downarrow$      & 37.39   &26.02 & 30.47 & 39.24   & \textbf{\cellcolor[HTML]{D7FFD7} 20.86} & 32.39      & 27.47   & 24.07   \\
& Acc. $\uparrow$       & 0.9498  & 0.9463 & 0.9548 & 0.9433  & 0.9584 & 0.9493    & 0.9587   & \textbf{\cellcolor[HTML]{D7FFD7} 0.9603}  \\
& IoU $\uparrow$        & 0.4969  & 0.5673 & 0.4891 & 0.4343  & 0.5762 & 0.4750   & 0.5599    & \textbf{\cellcolor[HTML]{D7FFD7} 0.6132} \\
\multirow{-5}{*}{\begin{tabular}[c]{@{}l@{}}BU\\ Images\end{tabular}} 
& F1 $\uparrow$         & 0.5951  & 0.6562 & 0.5882 & 0.5470   & 0.6613 & 0.5768       & 0.6516 & \textbf{\cellcolor[HTML]{D7FFD7} 0.7025} \\
\hline
\end{tabular} \vspace{-0.3cm}
\end{table}

\fcircle[fill=deblue]{2pt} \textbf{2D segmentation} In Table~\ref{tab:results-2D}, we provide a comparison of the performance of different segmentation methods across three datasets. We first compared with the baseline model U-Net~\cite{ronneberger2015u}, USODE~\cite{cheng2023continuous} and U-KAN~\cite{li2024u}. In addition, we compared with the U-KAN-like structure method UNeXt~\cite{valanarasu2022unext} and Rolling-Unet~\cite{liu2024rolling}.
We also compared our method with the well-known transformer method TransUNet~\cite{petit2021u} and state-of-the-art mamba like method MMLA-UNet~\cite{jiang2024mlla}. We report our performance in terms of Dice, HD95, Accuracy (Acc.), IoU, and F1 Score. Our proposed method consistently outperforms other segmentation models across multiple datasets, with improvements in key metrics such as {Dice score}, {HD95}, {accuracy}, and {F1 score}. On the {Kvasir-SEG} dataset, our method achieves a {Dice score of 0.8456}, representing an improvement of {14.6\% to 21.5\%} over U-KAN (0.7331) and USODE (0.7465), respectively. In terms of {HD95}, we achieve a {47.7\% reduction} (from 48.40 to 25.26), demonstrating superior boundary accuracy. We also show improvements in {accuracy} (up to {5.5\%} over MLLA-U-Net) and {F1 score} (41.8\% improvement over U-KAN). 
On {BU images}, our model achieves a {Dice score of 0.8397}, improving by {3.2\%} over U-KAN (0.8140), with the {F1 score} and {IoU} also showing improvements ranging from {7.8\% to 9.5\%}.

These results highlight the key strengths of our method: superior {segmentation precision} and \textbf{boundary delineation}. The significant improvements in {HD95}, particularly a \textbf{35.7\% reduction} over {U-Net}, show that our method better captures the fine details of the segmented regions. Additionally, the {efficiency} of our model, with {constant memory costs} and {GPU-based training}, ensures scalability, unlike traditional U-Net-based models that struggle with memory consumption and training time. Overall, our model demonstrates {superior performance} with \textbf{up to 40\% improvement} across multiple datasets, making it a \textbf{state-of-the-art solution} for medical image segmentation.

\begin{figure}[t!]
\centering
\includegraphics[width=1\textwidth]{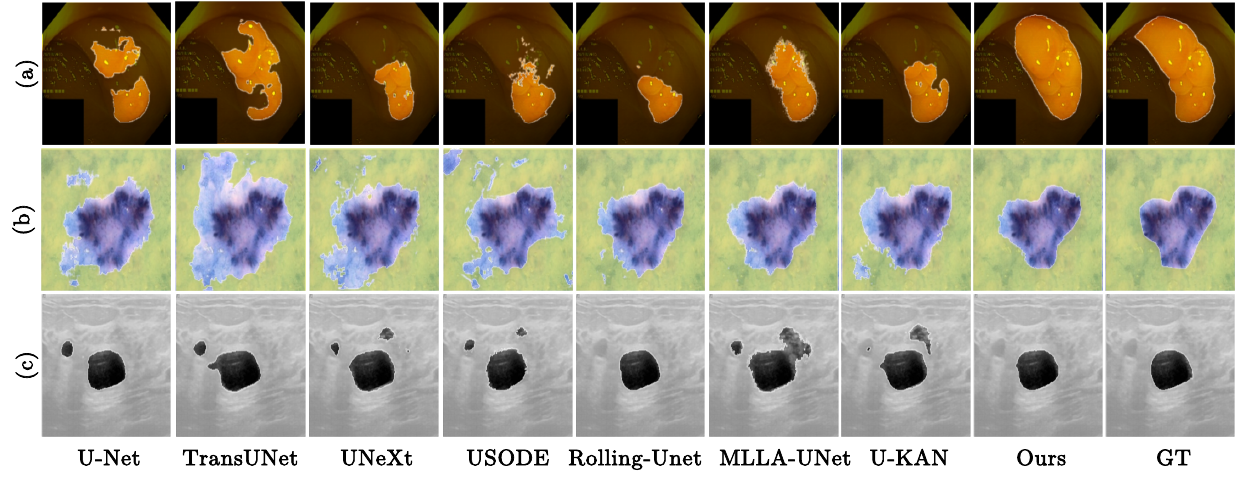}
\caption{The visualisation of the 2D segmentation task across the three datasets: (a) Kvasir-SEG~\cite{jha2020kvasir}, (b) ISIC Challenge~\cite{gutman2016skin}, and (c) Breast Ultrasound Images~\cite{al2020dataset}. 
}
\label{fig::2D}
\end{figure}

Figure~\ref{fig::2D} demonstrates that our method achieves cleaner, more precise segmentations closely matching GT labels across all datasets. While Rolling-U-Net performs well, it shows minor detail inaccuracies. Other methods are less accurate, often missing key structures or creating fragmented regions, underscoring the superior performance of Implicit U-KAN 2.0.\\
\begin{table}[t!]
\centering
\begin{minipage}{0.4\textwidth}
\centering
\caption{Comparison of 3D segmentation on Spleen dataset.
}
\label{tab::3D}
\begin{tabular}{>{\centering\arraybackslash}p{2.15cm} |>{\centering\arraybackslash}p{2.15cm} }
\hline
\cellcolor[HTML]{EFEFEF}\textsc{Method} & \cellcolor[HTML]{EFEFEF}Dice \\ \hline
\cellcolor[HTML]{FFFFFF} U-Net 3D & 0.9021 \\
\cellcolor[HTML]{FFFFFF} U-KAN 3D & 0.9591 \\
\cellcolor[HTML]{FFFFFF} Ours & \cellcolor[HTML]{D9FFD9}0.9687 \\ \hline
\end{tabular}
\label{diagonal}
\end{minipage}%
\hfill
\begin{minipage}{0.55\textwidth}
\centering
\caption{
Comparison between different noise levels in terms of Dice Socee
}\label{noise}
\begin{tabular}{>{\centering\arraybackslash}p{2.15cm} |>{\centering\arraybackslash}p{1.35cm} |>{\centering\arraybackslash}p{1.35cm} |>{\centering\arraybackslash}p{1.35cm} }
\hline
\cellcolor[HTML]{EFEFEF}\textsc{Method} & \multicolumn{3}{c}{\cellcolor[HTML]{EFEFEF}ISIC
Challenge} \\ \cline{2-4} 
\cellcolor[HTML]{EFEFEF}\textsc{} & \cellcolor[HTML]{EFEFEF}\textsc{0.0} & \cellcolor[HTML]{EFEFEF}\textsc{0.2} & \cellcolor[HTML]{EFEFEF}\textsc{0.4}  \\ \hline
\cellcolor[HTML]{FFFFFF} U-KAN & 0.9190 & 0.4064 & 0.4064  \\
\cellcolor[HTML]{FFFFFF} Ours & \cellcolor[HTML]{D9FFD9}0.9330 & \cellcolor[HTML]{D9FFD9}0.9225 & \cellcolor[HTML]{D9FFD9}0.9079\\  \hline
\end{tabular}
\end{minipage} \vspace{-0.4cm}
\end{table}
\fcircle[fill=deblue]{2pt} \textbf{3D segmentation.} We implemented both our method and U-KAN in 3D version. From the Table~\ref{tab::3D}, we observe that our proposed Implicit U-KAN 2.0 method significantly outperforms both U-Net 3D~\cite{cciccek20163d} and U-KAN 3D. Specifically, our method achieves a Dice score of 0.9687, which is notably higher than the 0.9021 achieved by U-Net 3D and also surpasses U-KAN 3D's score of 0.9591. This result underscores the superior segmentation performance of our proposed method across different dimensions, highlighting its effectiveness in achieving higher accuracy, as reflected in the Dice score. 

\fcircle[fill=deblue]{2pt} \textbf{Ablation Studies. }
In the comparison of segmentation performance under different noise levels on the ISIC Challenge dataset, our proposed Implicit U-KAN2.0 outperforms U-KAN at all noise levels. At noise level 0.2, our model achieves a Dice score of 0.9225, while U-KAN drops significantly to 0.4064, demonstrating a 126\% improvement. Even at the highest noise level (0.4), our method still delivers a Dice score of 0.9079, compared to U-KAN's 0.4064, showing a 123\% improvement. This highlights the advantage of our approach in handling noisy data, with continuous feature evolution via SONO providing smoother approximations and maintaining robust segmentation despite increased noise. Unlike traditional models, which struggle with noisy data, our model excels in delivering stable and accurate segmentations, making it particularly suitable for real-world clinical applications where image quality can be compromised.
\section{Conclusion}
We introduce implicit U-KAN 2.0. It leverages the SONO-Block and SONO-MultiKAN Block, achieving superior accuracy across three benchmark datasets. Our of second-order NODEs blocks enhances both efficiency and noise resistance, while the MultiKAN layer blocks improves interpretability. Additionally, our method outperforms both U-Net and U-KAN in 3D segmentation, highlighting its robustness  in handling complex medical image segmentation tasks.
\subsubsection{Acknowledgments} 
CWC and JAMZ are supported by the Swiss National Science Foundation under grant number 20HW-1 220785. YC is funded by an AstraZeneca studentship and a Google studentship. CBS  acknowledges support from the Philip Leverhulme Prize, the Royal Society Wolfson Fellowship, the EPSRC grants EP/S026045/1, EP/T003553/1, EP/N014588/1, the Wellcome Innovator Award RG98755 and the Alan Turing Institute.
AIAR gratefully acknowledges the support of the Yau Mathematical Sciences Center, Tsinghua University. This work is also supported by the Tsinghua University Dushi Program.

\subsubsection{Disclosure of Interests}
The authors declare that they have no competing interests.

\bibliographystyle{splncs04}
\bibliography{Paper-2894}

\end{document}